\documentclass[aps,prb,preprint]{revtex4-1}
\usepackage{siunitx}
\usepackage{amsmath}
\usepackage{bm}
\usepackage{subcaption}
\usepackage{graphicx}

\newcommand\numberthis{\addtocounter{equation}{1}\tag{\theequation}}

\begin{document}

\title{Extending solid-state calculations to ultra long-range length scales}

\author{T. M\"{u}ller}
\affiliation{Max-Planck-Institut f\"{u}r Mikrostrukturphysik, Weinberg 2,
 06120 Halle, Germany}
\author{S. Sharma}
\affiliation{Max-Born-Institut für Nichtlineare Optik und Kurzzeitspektroskopie,
 Max-Born-Strasse 2A, 12489 Berlin, Germany}
\author{E. K. U. Gross}
\affiliation{Fritz Haber Center for Molecular Dynamics, Institute of Chemistry,
The Hebrew University of Jerusalem, 91904 Jerusalem, Israel}
\author{J. K. Dewhurst}
\affiliation{Max-Planck-Institut f\"ur Mikrostrukturphysik, Weinberg 2,
 06120 Halle, Germany}
\email{dewhurst@mpi-halle.mpg.de}

\date{\today}

\begin{abstract}
We present a method which enables solid-state density
functional theory calculations to be applied to systems of almost unlimited size.
Computations of physical effects up to the micron length scale but
which nevertheless depend on the microscopic details of the electronic structure,
are made possible.
Our approach is based on a generalization of the Bloch state which involves an
additional sum over a finer grid in reciprocal space around each ${\bf k}$-point.
We show that this allows for modulations in the density and magnetization of arbitrary
length on top of a lattice-periodic solution. Based on this, we derive a set of
ultra long-range Kohn-Sham equations. We demonstrate our method with a sample
calculation of bulk LiF subjected to an arbitrary external potential
containing nearly 3500 atoms. We also confirm the accuracy of the method by
comparing the spin density wave state of bcc Cr against a direct super-cell calculation
starting from a random magnetization density. Furthermore, the spin spiral state of
$\gamma$-Fe is correctly reproduced and the screening by the density of a
saw-tooth potential over 20 unit cells of silicon is verified.
\end{abstract}

\maketitle

\section{Introduction}
Density functional theory (DFT) \cite{hohenberg_inhomogeneous_1964} has had a
tremendous impact on solid-state physics and is, due to its computational
efficiency, at the heart of modern computer based material research. Since its
original proposal, further developing DFT has been an ongoing process. Extensions
to DFT typically include extra densities in addition to the charge density, such
as the magnetization \cite{barth_1972}, current density \cite{vignale_1987} or
the superconducting order-parameter \cite{oliveira_1988}.
Another fundamental extension of DFT was the generalization to time-dependent
systems \cite{gross_1984} enabling accurate calculations of dynamical properties
of molecules and solids.
While these extensions allowed for a more in-depth understanding of microscopic
properties, not much progress has been made in applying DFT to effects in solids
occurring on larger, mesoscopic length scales.
Such effects include long-ranged quasiparticles, magnetic domains or spatially
dependent electric fields. As DFT is a formally exact theory, the underlying
physics for such phenomena are readily at hand, yet actual calculations remain
very difficult.
In a typical calculation, a single unit cell is solved with periodic boundary
conditions, thus effects extending far beyond the size of a single unit cell are
lost. While it is, in principle, possible to use ever larger super-cells, in
practice one quickly reaches the limit of computational viability. This is
mostly due to the poor scaling with the number of atoms,
$\sim \mathcal{O} ( N_{\rm atom}^3 )$, which plagues all
computer programs with a systematic basis set and limits calculations to systems
containing a maximum of $\sim 1000$ atoms.
Recent progress based on linear scaling approaches \cite{godecker_1999} was able
to increase the computable system size considerably.
Linear scaling approaches, however, require a ``nearsightedness'' of the system.
While this might be fulfilled for effects strictly related to the
charge density, this is certainly not fulfilled for large magnetic systems, such
as magnetic domains.

In this work we propose a fundamentally different approach to drastically extend the
length scale of DFT calculations without significantly increasing the
computational cost. 
Our approach relies on altered Bloch states and can be understood as a
generalization of the spin-spiral ansatz\cite{sandratskii_energy_1986}, which
emerges as a special case of our ansatz. In the spin-spiral ansatz, a
momentum-dependent phase is added to the normal Bloch state. It then becomes
possible to compute a large, extended spiraling magnetic moment with a
single unit cell. While this is computationally very efficient, it is, at the
same time, the biggest limitation of the spin-spiral ansatz: It allows only
for a change in the direction of the magnetization while the magnitude of the
magnetization and the charge density remain unaltered.
We overcome this limitation by introducing an additional sum in the Bloch
states over a finer grid in reciprocal space around each ${\bf k}$-point.
The resulting densities then become a Fourier series with a controllable
periodicity, which may extend far beyond the length scale of a single unit cell.

\section{Ultra long-range ansatz}

The systems we will focus on in this article are described by the Kohn-Sham (KS) Hamiltonian 
of spin-density
functional theory (atomic units are used throughout):
\begin{equation}\label{eq:sdft}
 \hat{H}_0 = - \frac{\nabla^2}{2} + v_s({\bf r})
 + {\bf B}_s({\bf r}) \cdot \bm{\sigma}.
\end{equation}
The KS potential $v_s({\bf r}) = v_{\rm ext}({\bf r}) +
v_{\rm H}({\bf r}) + v_{\rm xc}({\bf r})$
consists of an external potential $v_{\rm ext}$, a Hartree potential
$v_{\rm H}$ and an exchange-correlation (xc) potential $v_{\rm xc}$.
Similarly, the KS magnetic field ${\bf B}_s({\bf r}) =
\frac{1}{2c}{\bf B}_{\rm ext}({\bf r}) + {\bf B}_{\rm xc}({\bf r})$
can be decomposed into an external field ${\bf B}_{\rm ext}$ and an
xc-field ${\bf B}_{\rm xc}$.

We will start off by extending
the KS wave functions. From that we will derive altered charge and
magnetization densities. Finally we will derive a long-range Hamiltonian and the
matrix elements associated with it.

\subsection{Wave function and densities}
\begin{figure}
\centering
\begin{subfigure}{0.5\columnwidth}
  \centering
  \includegraphics[width=0.8\linewidth]{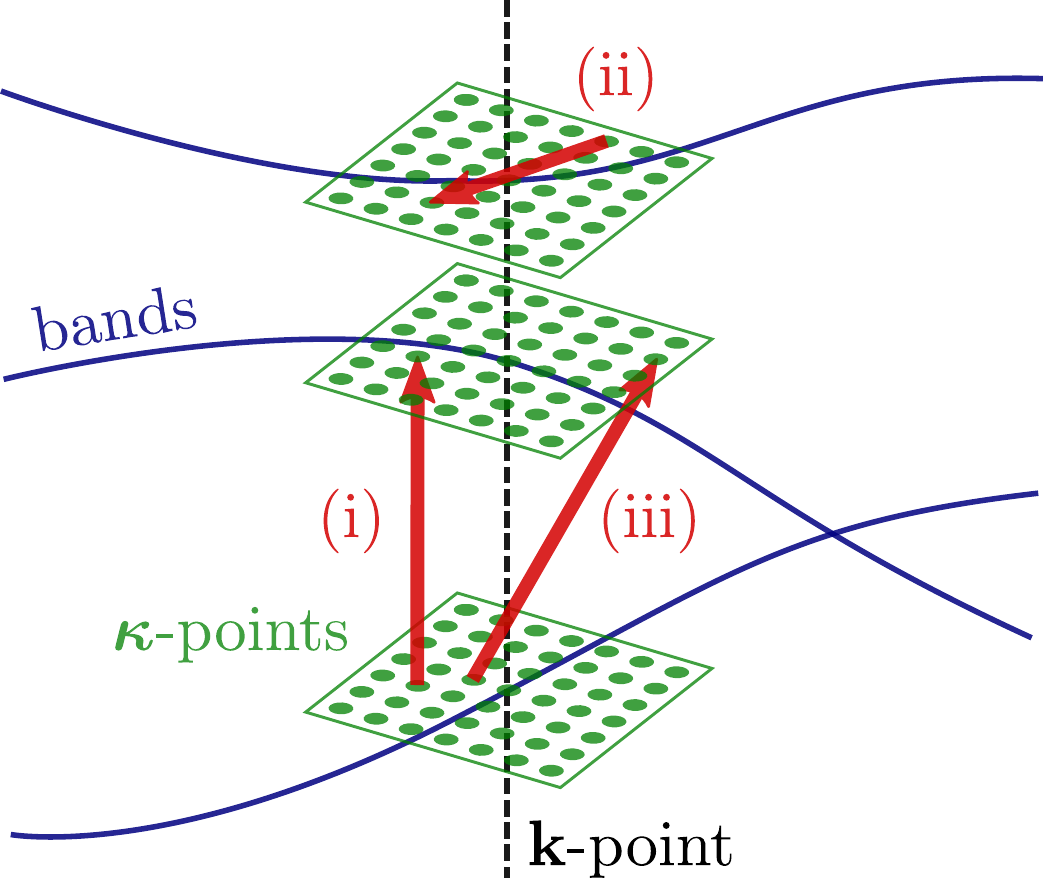}
  \caption{}
  \label{fig:kappa}
\end{subfigure}%
\begin{subfigure}{0.5\columnwidth}
  \centering
  \includegraphics[width=\linewidth]{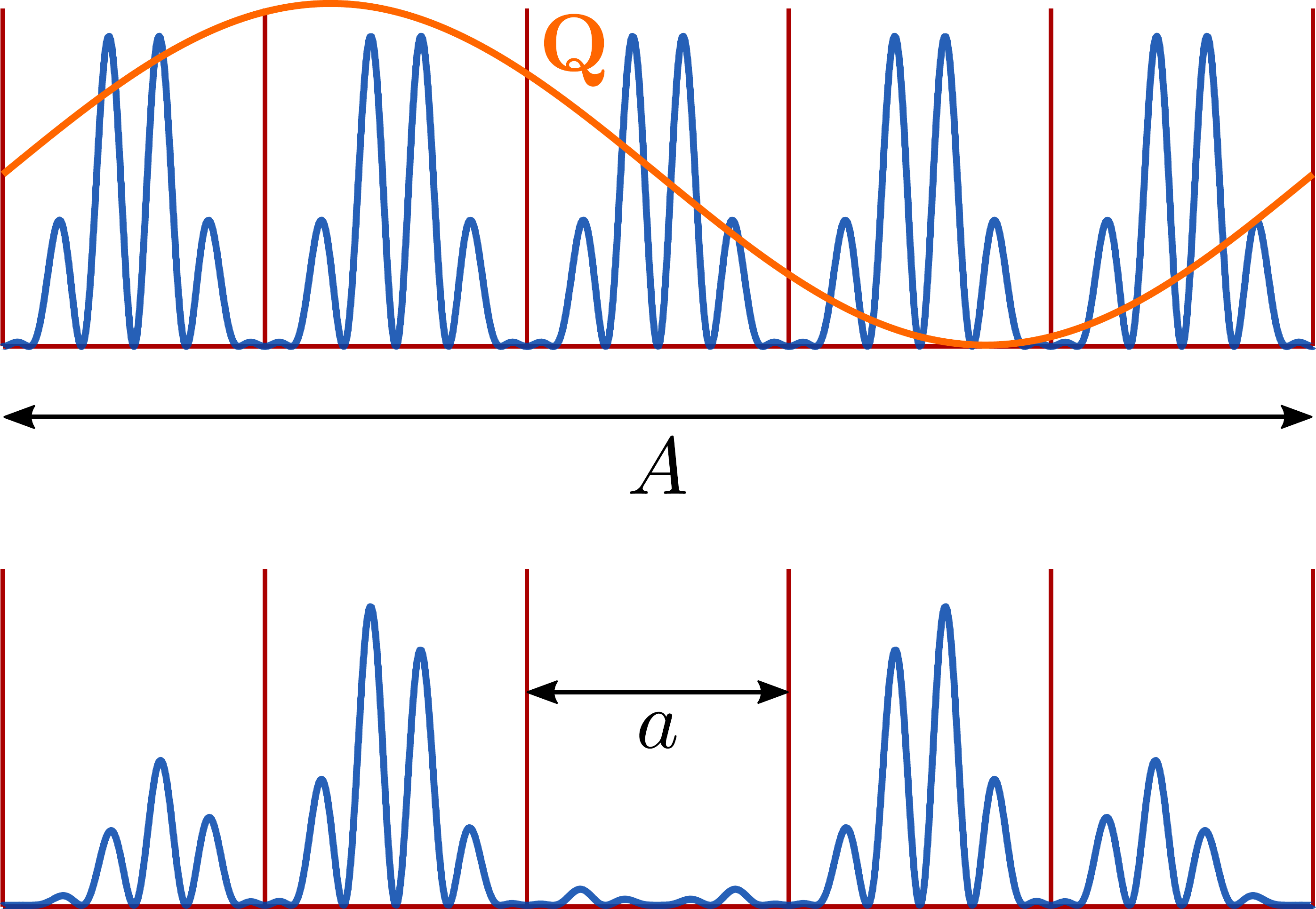}
  \caption{}
  \label{fig:mod}
\end{subfigure}
\caption{\raggedright (a) Schematic of the $\bm{\kappa}$-point grid. For each
${\bf k}$-point (black dashed line) all bands (blue) are augmented with a fine
grid of $\bm{\kappa}$-points (green). Three different types of couplings
between $\bm{\kappa}$-points corresponding to different length scales are
possible. (i) A coupling between two identical $\bm{\kappa}$-points but with
different band indices (ii) a coupling between different $\bm{\kappa}$-points
sharing the same band index, and (iii) a coupling between different
$\bm{\kappa}$-points with different band indices. The maximum length scale of
the calculation may be chosen by adjusting the $\bm{\kappa}$-point grid. (b) A
schematic of the long range approach. The red lines indicate unit cells.
The lattice periodic density $\rho_{{\bf Q}}$  (blue) is altered by a
${\bf Q}$-dependent modulation (orange) with a different periodicity. The result
(lower graph) depends on both, the long-range modulation and the lattice
periodic solution. $a$ is the lattice constant of a unit cell and $A$ is the
lattice constant of the ultracell, which is the smallest cell that
contains the full long-range solution.}
\end{figure}

Bloch states of the form
$\varphi_{i{\bf k}}({\bf r})=u_{i{\bf k}}({\bf r})e^{i {\bf k}\cdot {\bf r}}$,
where $u_{i{\bf k}}$ is a lattice-periodic spinor function,
are used in standard solid-state calculations.
The central idea of our approach is a generalization of this Bloch state to
include long-range fluctuations. A similar
idea was put forward with the spin-spiral ansatz\cite{sandratskii_energy_1986},
where a momentum-dependent phase is applied to the normal Bloch spinor state.
Our ultra long-range ansatz employs, in addition,  momentum-dependent expansion
coefficients which allow for changes in magnitude of the densities
from cell to cell. For a fixed ${\bf k}$-vector our new Bloch-like state reads:
\begin{align}\label{eq:lBloch}
 \Phi_{\alpha}^{{\bf k}}({\bf r}) &= \frac{1}{\sqrt{N_u}}
 \sum_{i \bm{\kappa}}c^{\alpha}_{i {\bf k}+\bm{\kappa}} 
 \begin{pmatrix}
 u_{i{\bf k}}^\uparrow({\bf r}) \\
 u_{i{\bf k}}^\downarrow({\bf r})
 \end{pmatrix}
 e^{i({\bf k}+\bm{\kappa})\cdot{\bf r}}
\end{align}
where $u_{i {\bf k}}^{\uparrow\downarrow}$ are the normalized
orbitals of a lattice-periodic system,
$i$ is a band index and ${\bf k}$ a reciprocal space vector,
$c^{\alpha}_{i {\bf k}+\bm{\kappa}}$ are complex coefficients to be determined
variationally (or by propagating in time) and $\alpha$ labels a particular
long-range state. The vectors $\bm{\kappa}$ live on a finer grid
around each ${\bf k}$-point in reciprocal space (Fig. \ref{fig:kappa}), which we
use to sample long-range effects. Finally $N_u$ is a normalization factor which
is equal to the number of unit cells on which $\Phi_\alpha^{{\bf k}}$ is periodic.
Note that we have used the lattice periodic parts of the orbitals at ${\bf k}$
and not ${\bf k}+\bm{\kappa}$. In principle, both are complete basis sets
capable of expanding any lattice-periodic function. In practice, the choice of
using $u_{i{\bf k}}^{\uparrow\downarrow}$ over
$u^{\uparrow\downarrow}_{i{\bf k}+\bm{\kappa}}$
is more efficient for determining the density, magnetization and
Hamiltonian matrix elements.

From this wave function, we can construct a charge and magnetization density:
\begin{equation}\label{eq:Cdens}
 \rho({\bf r}) = \frac{1}{N_k} \sum_{{\bf k},\alpha}
 f_{\alpha}^{{\bf k}} \Phi_{\alpha}^{{\bf k} \dagger}({\bf r})
 \Phi_{\alpha}^{{\bf k}}({\bf r})
\end{equation}
\begin{equation}\label{eq:Mdens}
 {\bf m}({\bf r}) = \frac{1}{N_k} \sum_{{\bf k}, \alpha}
 f_{\alpha}^{{\bf k}} \Phi_{\alpha}^{{\bf k} \dagger}({\bf r})
 \bm{\sigma} \Phi_{\alpha}^{{\bf k}}({\bf r})
\end{equation}
with the number of ${\bf k}$-points $N_{{\bf k}}$ and the ultra long-range
occupation numbers $f_\alpha^{{\bf k}}$ associated with the orbitals
$\Phi_{\alpha}^{{\bf k}}$.
The charge and magnetization density obtained from this wave function take the form
\begin{align}\label{eq:fdens}
\begin{split}
 \rho({\bf r}) &= \sum_{{\bf Q}} \rho_{{\bf Q}}({\bf r})
 e^{i {\bf Q} \cdot {\bf r}}, \\
 {\bf m}({\bf r}) &= \sum_{\bf Q}
 {\bf m}_{\bf Q}({\bf r}) e^{i {\bf Q} \cdot {\bf r}}
\end{split}
\end{align}
with ${\bf Q} = \bm{\kappa} - \bm{\kappa'}$. The partial densities
$\rho_{{\bf Q}}$ and ${\bf m}_{{\bf Q}}$ in Eq. (\ref{eq:fdens}) are complex in general
and act as lattice-periodic Fourier coefficients. The resulting real-space densities
$\rho({\bf r})$ and ${\bf m}({\bf r})$ are real functions, which, depending
on the values of ${\bf Q}$,
will have a periodicity larger than the length scale of a unit cell
(Fig. \ref{fig:mod}). By adjusting the underlying $\bm{\kappa}$-lattice, it is
therefore possible to change the ${\bf Q}$-vectors and hence allow for variations
of arbitrary length in the system. The ${\bf Q}=0$ term deserves special mention,
as it corresponds to the full lattice-periodic solution. We emphasize that there
is no restriction on the magnitude of $\rho_{{\bf Q}}, {\bf m}_{{\bf Q}}$ and we
are thus able to expand arbitrary modulations in the charge and magnetization
densities. This is a key difference compared to the spin-spiral ansatz
\cite{sandratskii_energy_1986}.

The Fourier coefficients $\rho_{{\bf Q}}$ and ${\bf m}_{{\bf Q}}$ can be calculated
efficiently by first calculating the wave function in Eq. (\ref{eq:lBloch})
for a subset of unit cells given by a set of real-space lattice vectors
$\{{\bf R}_i\}$.
We choose the ${\bf R}_i$-vectors to be the conjugate real-space vectors of the
${\bf Q}$-vectors. The wave function in a single unit cell is then given by a
sum over $n$ and a fast Fourier transform in $\bm{\kappa}$ of the
coefficients $c^{\alpha}_{n {\bf k}+\bm{\kappa}}$:
\begin{equation}\label{eq:waveR}
 \Phi_{\alpha}^{{\bf k}}({\bf r}+{\bf R}_i)\approx
 e^{i {\bf k} \cdot ({\bf r}+{\bf R}_i)}\sum_{n}
 \begin{pmatrix}
 u_{n {\bf k}}^\uparrow({\bf r}) \\
 u_{n {\bf k}}^\downarrow({\bf r})
 \end{pmatrix}
 \sum_{\bm{\kappa}}
 c^{\alpha}_{n {\bf k}+\bm{\kappa}}
 e^{i \bm{\kappa} \cdot {\bf R}_i}
\end{equation}
where ${\bf r}$ is restricted to a single unit cell and we have assumed that
$|\bm{\kappa}\cdot{\bf r}|\ll 1$. Note also that the normalization constant
$1/\sqrt{N_u}$ has been removed. This ensures that observables such as
charge and energy are calculated per unit cell rather than per ultracell.
From this, we compute the
charge and magnetization densities on the same grid,
i.e. $\rho_{i} = \rho({\bf r} + {\bf R}_i)$
and ${\bf m}_i = {\bf m}({\bf r} + {\bf R}_i)$.
This set can then be partially (fast) Fourier transformed to reciprocal space to
obtain $\rho_{{\bf Q}}({\bf r})$ and
${\bf m}_{{\bf Q}}({\bf r})$:
\begin{align}\label{eq:coeff}
\begin{split}
 \rho_{\bf Q}({\bf r}) &= \frac{1}{N_{{\bf R}}}
 \sum_{i} \rho({\bf r} + {\bf R}_i)\,e^{-i {\bf Q} \cdot {\bf R}_i} \\
 {\bf m}_{\bf Q}({\bf r}) &= \frac{1}{N_{{\bf R}}} \sum_{i} {\bf m}
 ({\bf r} + {\bf R}_i)\,e^{-i {\bf Q} \cdot {\bf R}_i}
\end{split}
\end{align}
Here $N_{{\bf R}}$ denotes the number of ${\bf R}$-vectors chosen. 
With the densities at hand, we will now focus on generalizing the Hamiltonian
such that meaningful, non-trivial values for the expansion coefficients
$c^\alpha_{n{\bf k}+\bm{\kappa}}$ in Eq. (\ref{eq:lBloch}) are obtained.

\subsection{Long-range Hamiltonian}
The ultra long-range Hamiltonian retains the full lattice periodic KS
Hamiltonian $\hat{H}_0$ given in Eq. (\ref{eq:sdft}), but also has an
additional ``modulation'' term
\begin{equation}\label{eq:altHam}
 \hat{H} = \hat{H}_{0} + \sum_{{\bf Q}} \hat{H}_{{\bf Q}}({\bf r})
 e^{i {\bf Q} \cdot {\bf r}}.
\end{equation}
The total Hamiltonian $\hat{H}$ is thus decomposed in the same way as the
charge and magnetization densities in Eq. (\ref{eq:fdens}).
For a KS system like Eq. (\ref{eq:sdft}), our ``modulation'' Hamiltonian reads
\begin{equation}\label{eq:longHam}
 \hat{H}_{{\bf Q}}\left( {\bf r} \right) =
 V_{{\bf Q}}({\bf r})+{\bf B}_{{\bf Q}}({\bf r}) \cdot \bm{\sigma},
\end{equation}
where $V_{{\bf Q}}({\bf r})$ and ${\bf B}_{{\bf Q}}({\bf r})$
are again complex, lattice periodic Fourier coefficients
and contribute to long-ranged versions of the scalar potential and the magnetic
field, respectively. In the following we will discuss
these coefficients and how to compute them in more detail. We will start with
the scalar potential, which can again be decomposed into an external
potential $V_{{\bf Q}}^{\rm ext}({\bf r})$, a Hartree potential
$V^{\rm H}_{{\bf Q}}({\bf r})$ and an xc-potential $V^{\rm xc}_{{\bf Q}}({\bf r})$.
The coefficients $V_{{\bf Q}}^{\rm ext}({\bf r})$ of an external,
long-ranged potential can be freely chosen.
The coefficients for the long-ranged Hartree potential
$V^{\rm H}_{{\bf Q}}({\bf r})$ are obtained from the long-range density
in Eq. (\ref{eq:fdens}):
\begin{equation}\label{eq:VhQ}
 V^{\rm H}_{{\bf Q}}({\bf r}) =
 \int d^3 r' \frac{\rho_{{\bf Q}}({\bf r'})}{\left|{\bf r}-{\bf r'}\right|}
 e^{-i {\bf Q}\cdot({\bf r} - {\bf r}')}.
\end{equation}
This may be performed efficiently by further Fourier transforming
$\rho_{{\bf Q}}({\bf r})$ to $\rho_{{\bf Q}}({\bf G})$ where ${\bf G}$ is a
reciprocal lattice vector. The Hartree potential is then determined directly via
$V^{\rm H}_{{\bf Q}}({\bf G})=4\pi\rho_{{\bf Q}}({\bf G})/|{\bf G}+{\bf Q}|^2$
and can be subsequently Fourier transformed back to real-space.
This is easily extended to the case of the augmented plane wave basis by using
the method of Weinert \cite{weinert_solution_1981}.

Next we will determine the coefficients associated with the xc-interaction. An
important difference compared to the Hartree potential is that the
xc-functional is inherently non-linear, therefore the naive approach
$V_{{\bf Q}}^{\rm xc} = V_{\rm xc}[\rho_{{\bf Q}}]$
may introduce a mixing of the real and imaginary part of $\rho_{{\bf Q}}$. 
Instead we first Fourier transform the density to real-space,
$\rho_{{\bf R}_i}({\bf r})$, and then
evaluate the xc-potential separately for each ${\bf R}$-vector. The
inverse Fourier transform is then applied to obtain
\begin{equation}\label{eq:VxcQ}
 V_{{\bf Q}}^{\rm xc}({\bf r}) =
 \frac{1}{N_{{\bf R}}} \sum_{i}
 V_{\rm xc}\left[\rho_{ {\bf R}_i}\right]({\bf r})
 e^{-i {\bf Q} \cdot {\bf R}_i}.
\end{equation}
It is worth noting that defining the long-range xc-functional this way does not
change how local an xc-functional inherently is, it is merely a Fourier
interpolation.

The magnetic field ${\bf B}_{{\bf Q}}$ in Eq. (\ref{eq:longHam}) consists of an
external field, an xc-field and a dipole-dipole field:
\begin{equation}\label{eq:BQ}
 {\bf B}_{{\bf Q}}({\bf r}) = \frac{1}{2c}{\bf B}_{{\bf Q}}^{\rm ext}({\bf r})
 + {\bf B}_{{\bf Q}}^{\rm xc}({\bf r})
 +\frac{1}{2c}{\bf B}_{{\bf Q}}^{\rm D}({\bf r}).
\end{equation}
Again, the external magnetic field may be chosen arbitrarily and the xc-field
can be computed analogously to the xc-potential
\begin{equation}\label{eq:BxcQ}
 {\bf B}_{{\bf Q}}^{\rm xc} \left({\bf r} \right) =
 \frac{1}{N_{{\bf R}}} \sum_{i} {\bf B}_{\rm xc}
 \left[\rho_{{\bf R}_i}, {\bf m}_{{\bf R}_i}\right]({\bf r})
 e^{-i {\bf Q}\cdot{\bf R}_i}.
\end{equation}
The last term in Eq. (\ref{eq:BQ}) corresponds to the magnetic field associated
with the magnetostatic dipole-dipole interaction
\begin{equation}\label{eq:Bdip}
 {\bf B}_{{\bf Q}}^{\rm D} \left({\bf r} \right) =
 \frac{1}{2c}\int d^3 r'\,\frac{3\,{\bf e}_{{\bf r}-{\bf r'}}
 \left({\bf m}_{{\bf Q}}({\bf r'}) \cdot
 {\bf e}_{{\bf r}-{\bf r'}} \right) - {\bf m}_{{\bf Q}}
 ({\bf r'})}{\left|{\bf r} - {\bf r}' \right|^3}
 e^{-i {\bf Q}\cdot({\bf r} - {\bf r}')},
\end{equation}
where ${\bf e}_{{\bf r}-{\bf r'}}$ is the unit vector along the direction
${\bf r}-{\bf r'}$. The contribution of the dipole-dipole interaction is
typically neglected in DFT calculations as it is usually small in comparison
with ${\bf B}^{\rm xc}$, which originates from the Coulomb exchange
interaction. As the coulomb exchange interaction is inherently short ranged,
the magnetic dipole-dipole interaction is expected to have a significant
contribution at larger length scales. We therefore include this term in
the ``modulation'' Hamiltonian.
 The derivation of a truly non-local, ${\bf Q}$-dependent
xc-potential is beyond the scope of this article but has been addressed
by Pellegrini, {\it et. al}\cite{PhysRevB.101.144401} for the
dipole interaction.

We conclude this section with a remark on the kinetic energy. The kinetic
energy operator ${\bf \hat{p}}^2/2$ does not explicitly depend on the
periodicity of the problem at hand. As the kinetic energy operator is already
included in $\hat{H}_{0}$ (Eqs. (\ref{eq:sdft}), (\ref{eq:altHam})), it
should not be added to $\hat{H}_{{\bf Q}}$.
It is important to note, however, that the kinetic energy is sensitive to
the shifts in reciprocal space of the wave function
(Eq. (\ref{eq:lBloch})) ${\bf k} \rightarrow {\bf k} + \bm{\kappa}$ which
should be taken into account.

\subsection{Hamiltonian Matrix Elements}
We will now focus on diagonalizing the long-range Hamiltonian in
Eq. (\ref{eq:altHam}).
For that we compute the matrix elements for a fixed ${\bf k}$-point in the
orbital basis of Eq. (\ref{eq:lBloch}) to evaluate
\begin{equation}\label{eq:matEl}
 \left\langle \varphi_{i {\bf k}+\bm{\kappa}} \right| \hat{H}_{0} +
 \hat{H}_{{\bf Q}} \left| \varphi_{j {\bf k}+\bm{\kappa'}}
 \right\rangle = \delta_{\bm{\kappa},\bm{\kappa'}}
 \left(O_{{\bf k}+\bm{\kappa},{\bf k}}^{\dag}\epsilon_{{\bf k}+\bm{\kappa}}^{0}
 O_{{\bf k}+\bm{\kappa},{\bf k}}\right)_{ij}
 + \left\langle \varphi_{i {\bf k}+\bm{\kappa}} \right|
 \hat{H}_{{\bf Q}} \left| \varphi_{j{\bf k}+\bm{\kappa'}} \right\rangle.
\end{equation}
Here ${\bf Q}=\bm{\kappa}-\bm{\kappa'}$,
$\epsilon_{{\bf k}+\bm{\kappa}}^{0}$ is the diagonal matrix of
eigenvalues of $\hat{H}_{0}$ at ${\bf k}+\bm{\kappa}$ and
$O_{{\bf k}+\bm{\kappa},{\bf k}}$ is the unitary overlap matrix between
the orbitals at ${\bf k}$ and ${\bf k}+\bm{\kappa}$, i.e.
\begin{equation}
 \left(O_{{\bf k}+\bm{\kappa},{\bf k}}\right)_{ij}=
 \sum_{s} \int d^3 r \, \varphi_{i s {\bf k}+\bm{\kappa}}^*({\bf r})
 \exp(i\bm{\kappa}\cdot{\bf r})\varphi_{j s {\bf k}}({\bf r}).
\end{equation}
This overlap matrix is required because our chosen basis is the
set of orbitals at ${\bf k}$ and not those at ${\bf k}+\bm{\kappa}$.
The overlap matrix $O_{{\bf k}+\bm{\kappa},{\bf k}}$ may however not be
strictly unitary in practice. This may be because of numerical inaccuracies
but also because the basis is finite and there could be bands of a
particular character at some $({\bf k}+\bm{\kappa})$-points but not at others.
Unitarity is necessary for preserving the eigenvalues
$\epsilon_{{\bf k}+\bm{\kappa}}^{0}$ and we ensure this
by first performing a singular value decomposition
$O_{{\bf k}+\bm{\kappa},{\bf k}}=U\Sigma T^{\dag}$ and then making the substitution
$O_{{\bf k}+\bm{\kappa},{\bf k}}\rightarrow UT^{\dag}$. One can show
that this new matrix is the closest (in the sense of the Frobenius norm)
unitary matrix to the original.

What remains to be done is the calculation of the matrix elements of
$\hat{H}_{{\bf Q}}$ in Eq. (\ref{eq:longHam}). We start with the the scalar
potential and find:
\begin{align*}\label{eq:matSC}
 \left\langle \varphi_{n {\bf k}}^{\bm{\kappa}} \right|\hat{V}_{{\bf Q}} \left|
 \varphi_{n'{\bf k} }^{\bm{\kappa'}} \right\rangle
 = \sum_{s} \int_{\rm unit} d^3 r \, u_{n s {\bf k}}^*({\bf r})
 u_{n' s {\bf k}}({\bf r}) V_{{\bf Q}} \left( {\bf r} \right),\numberthis
\end{align*}
where $s=\uparrow,\downarrow$ is a spin index.  Here we have defined
$\varphi_{n {\bf k}}^{\bm{\kappa}}({\bf r})\equiv
 \varphi_{n {\bf k}}({\bf r})e^{i \bm{\kappa} \cdot {\bf r}}$. In the first step we converted the
integral over the ultracell into an integral over a unit cell and a sum over
all unit cells in the ultracell
$\int_{\rm ultra} d^3 r \to \sum_{{\bf R}_u}\int_{\rm unit} d^3 r$
and made use of the lattice periodicity of $u_{n{\bf k}}({\bf r})$.
In the second step we then carried out the sum over ${\bf R}_u$ followed by the
sum over ${\bf Q}$. The matrix elements for the ultracell can thus be expressed
by a simple unit cell integration.
Similarly we find for the magnetic field contribution:
\begin{align}
\begin{split}
 \left\langle \varphi_{n {\bf k}}^{\bm{\kappa}}  \right|
 \hat{\bf B}_{\bf Q} \cdot \bm{\sigma}
 \left| \varphi_{n'{\bf k} }^{\bm{\kappa'}} \right\rangle =
 \int_{\rm unit} d^3 r \,&
  u_{\uparrow n {\bf k}}^*({\bf r}) u_{\downarrow n' {\bf k}}({\bf r})
 \left(B_{\bf Q}^x({\bf r}) - i B_{\bf Q}^y({\bf r}) \right) \\
 +&u_{\downarrow n{\bf k}}^*({\bf r}) u_{\uparrow n'{\bf k}}({\bf r})
  \left(B_{\bf Q}^x({\bf r}) + i B_{\bf Q}^y({\bf r}) \right) \\
 +&\left(u_{\uparrow n{\bf k}}^*({\bf r}) u_{\uparrow n'{\bf k}}({\bf r})
 -u_{\downarrow n{\bf k}}^*({\bf r}) u_{\downarrow n'{\bf k}}({\bf r})\right)
 B_{\bf Q}^z({\bf r}).
\end{split}
\end{align}

\section{Numerical implementation}
In this section we will address how to implement the ultra long-range ansatz in
practice. The discussions in this section are based on our implementation in the
Elk electronic structure code \cite{elk}, which is an all-electron code using
the full potential linearized augmented plane wave (FP-LAPW) method.

\subsection{Self-consistent solution}
$\hat{H}$ in Eq. (\ref{eq:altHam}) is a KS system in which the potentials are
functionals of the partial densities
$\rho_{\bf Q}({\bf r})$ and ${\bf m}_{\bf Q}({\bf r})$ in Eq. (\ref{eq:coeff}), which
in turn depend on the orbitals $\Phi_\alpha^{\bf k}({\bf r})$ from Eq. (\ref{eq:lBloch}).
Equation (\ref{eq:altHam}) thus needs to be solved self-consistently. We employ an
iteration scheme as it is usually done when solving KS systems:

{\raggedleft
\fbox{\parbox{0.98\textwidth}{
\begin{enumerate}
\item Solve the lattice periodic ground state, Eq. (\ref{eq:sdft}) and obtain
 the spinor orbitals
 $\begin{pmatrix}
  u_{n{\bf k}}^\uparrow({\bf r}) \\
  u_{n{\bf k}}^\downarrow({\bf r})
  \end{pmatrix}$
 as well as all Eigen energies $\epsilon^0_{n{\bf k}+\bm{\kappa}}$ associated
 with the ${\bf k}+\bm{\kappa}$-points.
\item Initialize the external long-range potentials via
 $V_{\bf Q}$ and ${\bf B}_{\bf Q}$ and the occupation numbers $f^{\bf k}_\alpha$.
\item
\begin{enumerate}
\item Compute the matrix elements of $\hat{H}_{\bf Q}$ in Eq. (\ref{eq:longHam}).
 Diagonalize $\hat{H}$ in Eq. (\ref{eq:altHam}) to obtain the expansion
 coefficients $c^\alpha_{n{\bf k}+\bm{\kappa}}$ as well as the long-range
 Eigen-energies $\epsilon^{\bf k}_\alpha$ for each ${\bf k}$-point.
\item Concurrently with the step above, accumulate the long-range densities
 $\rho_{\bf Q}({\bf r})$ and ${\bf m}_{\bf Q}({\bf r})$ from
 Eqs. (\ref{eq:Cdens}) and (\ref{eq:Mdens}).
 This is performed most efficiently by first calculating the long-range
 orbitals explicitly in real-space: $\Phi_{\alpha}^{{\bf k}}({\bf r}+{\bf R}_i)$.
\end{enumerate}
\item Calculate the new occupation numbers $f^{\bf k}_\alpha$.
\item Calculate new long-range potentials $V_{\bf Q}'$ and ${\bf B}_{\bf Q}'$.
 Mix the new potentials with the potentials from the previous iteration. Monitor
 the relative change in the potentials.
\item Repeat steps 3 to 5 until the change in the potentials is sufficiently small.
\end{enumerate}}}
}

We will discuss two steps in this self-consistent cycle in more detail.

First we will explain the order of calculating the energies
$\epsilon^{\bf k}_\alpha$ first, the densities
$\rho_{\bf Q}({\bf r})$ and ${\bf m}_{\bf Q}({\bf r})$
second and the occupation numbers $f_\alpha^{\bf k}$ third.
This seems counter-intuitive, as the densities depend on the occupation
numbers (Eqs. (\ref{eq:Cdens}) and (\ref{eq:Mdens})). However, as we are
performing a self-consistent cycle, the occupation numbers will converge to
the correct value as self-consistency is achieved.
Computing the occupation numbers last enables us to parallelize step 3
over the ${\bf k}$-point set in a single loop: For each ${\bf k}$-point, we
diagonalize
$\hat{H}^{\bf k} = \hat{H}^{\bf k}_0 + \hat{H}^{\bf k}_{\bf Q}$
and simultaneously compute
$\rho^{\bf k}_{\bf Q}({\bf r})$ and ${\bf m}^{\bf k}_{\bf Q}({\bf r})$.
These are added to the total density and magnetization.
The central computational gain is that this ordering is much less demanding
when it comes to memory: the coefficients $c^\alpha_{n{\bf k}+\bm{\kappa}}$ do
not have to be stored but can be calculated and used on-the-fly instead.

Second, we note that some care has to be taken during the mixing. We choose
to mix the complex Fourier coefficients
$V_{\bf Q}({\bf r})$ and ${\bf B}_{\bf Q}({\bf r})$ rather than their
real-space counterparts.
We also want to emphasize that in a typical calculation a rather slow mixing
should be applied. The Coulomb interaction in a large system will react very
strongly to any external perturbation because of the divergence of $1/Q^2$.
This can lead to substantial charge sloshing during convergence necessitating the
use of a small mixing parameter. This is an aspect of the method which
would benefit from further investigation and improvement. One possibility is
to use a screened Coulomb interaction to remove the divergence. This screening
could be slowly reduced to zero during the self-consistent loop to improve
the rate of convergence.

\subsection{${\bf k}$-point grids}
The underlying grids have to be chosen carefully in order to avoid computational
artifacts and to achieve a most efficient calculation. Ideally, the
smallest distance between ${\bf k}$-points should be
greater than the largest distance between $\bm{\kappa}$-points, i.e.
$\left|\bm{\kappa}-\bm{\kappa'}\right| < \left|{\bf k}-{\bf k'} \right|$.
This will ensure that the set ${\bf k}+\bm{\kappa}$ does not overlap for any
two ${\bf k}$-points, which may lead to double counting and an over-complete
basis set. Physically speaking, the length scales in the system should be well
separated, i.e. the modulation should be far larger than the size of a unit cell.
If $\left|\bm{\kappa}-\bm{\kappa'}\right|\approx \left|{\bf k}-{\bf k}'\right|$,
however, the system tends to have a size which can and should be solved with a
super-cell instead.

Many Fourier transformations need to be carried
out during each self-consistent step: with $e^{-i\bm{\kappa}\cdot{\bf R}_i}$
when calculating the wave function in Eq. (\ref{eq:waveR}),
and with $e^{-i{\bf Q}\cdot{\bf R}}$ when
calculating the densities in Eqs. (\ref{eq:Cdens}) and (\ref{eq:Mdens}); and the
xc-potential and -field, Eqs. (\ref{eq:VxcQ}) and (\ref{eq:BxcQ}).
This constitutes a major part of the computational effort and
it is therefore highly beneficial to carry out
all Fourier transformations via a Fast Fourier Transform (FFT). This requires
the underlying grid to be FFT compatible (having, in our case, radices 2, 3, 5 and 7).
Owing to ${\bf Q}=\bm{\kappa}-\bm{\kappa'}$ the ${\bf Q}$-point and
the $\bm{\kappa}$-point grids are dependent on one another. The number of
${\bf Q}$-points $n_{\bf Q}$ along a given direction $i$ is
$n_{\bf Q}^i = 2 n_{\bm{\kappa}}^i-1$. 
In our implementation, we ensure that the input ${\bf Q}$-grid snaps to the next
FFT compatible grid. We then choose the $\bm{\kappa}$-point grid such that
$2 n_{\bm{\kappa}}^i-1\le n^i_{\bf Q}$. This grid choice can sometimes result in
unmatched ${\bf Q}$-points, e.g. if $n_{\bf Q}=20$ and $n_{\bm{\kappa}}=10$
then the ${\bf Q}$-vectors are not symmetric around zero. While the unmatched
${\bf Q}$-point is ``dead-weight'' and remains zero throughout the calculation,
the speed up obtained by using a FFT outweighs having additional ${\bf Q}$-points.

\subsection{Computation of the Hartree and dipole interaction}

We will briefly address how to calculate the complex integrals appearing in the
scalar potential, Eq. (\ref{eq:VhQ}), and the magnetic field, Eq. (\ref{eq:Bdip}).
When computing the Hartree-potential in Eq. (\ref{eq:VhQ}), we solve Poisson's
equation using the method by Weinert\cite{weinert_solution_1981}, which can be
generalized to complex densities relatively easily.

The dipole interaction, Eq. (\ref{eq:Bdip}), can be solved for in a similar way
by evaluating Poisson's equation component-wise for the vector potential. From
classical electrodynamics, the vector potential associated with a magnetization
is given by:
\begin{equation}\label{eq:Adip}
 {\bf A}^{\rm dip}({\bf r})=
 \frac{1}{c} \int d^3r' \frac{\nabla\times{\bf m}\left({\bf r'}\right)}
 {\left|{\bf r}-{\bf r'}\right|}
\end{equation}
We partially Fourier transform both sides and obtain:
\begin{equation}
 \sum_{\bf Q}{\bf A}^{\rm dip}_{\bf Q} \left({\bf r}\right)
 e^{i{\bf Q}\cdot{\bf r}} =
 \frac{1}{c}\int d^3r' \, \frac{\nabla\times
 \sum_{\bf Q}{\bf m}_{\bf Q}({\bf r'})
 e^{i{\bf Q}\cdot{\bf r}'}}{\left|{\bf r}-{\bf r}'\right|}.
\end{equation}
We thus obtain for the coefficients of the vector potential:
\begin{equation}
 A_{{\bf Q},j}^{\rm dip}({\bf r}) = \frac{1}{c} \sum_{kl}
 \epsilon_{jkl} \int d^3r'\,
 e^{-i{\bf Q}\cdot\left({\bf r}-{\bf r}'\right)} \frac{\partial_k
 m_{{\bf Q},l}({\bf r}') + i Q_k m_{{\bf Q},l}({\bf r}')}
 {\left|{\bf r}-{\bf r}'\right|}
\end{equation}
Here $j,k,l$ indicate vector components and $\epsilon_{jkl}$ is the
Levi-Civita symbol. The coefficients
${\bf A}_{\bf Q}^{\rm dip}({\bf r})$ now have the same form as the Hartree
potential, Eq. (\ref{eq:VhQ}), and can also be computed by a complex version of
Weinert's method \cite{weinert_solution_1981}.
From this it is easy to obtain the magnetic field of the dipole interaction via
${\bf B}^{\rm dip}({\bf r})=\nabla \times {\bf A}^{\rm dip}({\bf r})$.
We find for the coefficients:
\begin{equation}
 B_{{\bf Q},j}^{\rm dip} = \sum_{kl}
 \epsilon_{jkl} \left[ \partial_k A_{{\bf Q},l}^{\rm dip}({\bf r})
 + i Q_k A_{{\bf Q},l}^{\rm dip}({\bf r}) \right].
\end{equation}
We point out that if we were to consider an exact theory for the current
density, the dipole vector potential, Eq. (\ref{eq:Adip}), should be
included in the Hamiltonian, corresponding to a Lorentz force generated by the
dipole-dipole interaction. 

\section{Results}

Three calculations for which the ultracell is
small enough to be amenable to super-cell calculations so that a detailed comparison is possible. We also performed
a calculation which would be considered too large to be treated as a super-cell.

\subsection{Spin-spirals in $\gamma$-Fe}

\begin{figure}[ht]
\centering
\begin{subfigure}{0.44\columnwidth}
 \includegraphics[width=\linewidth]{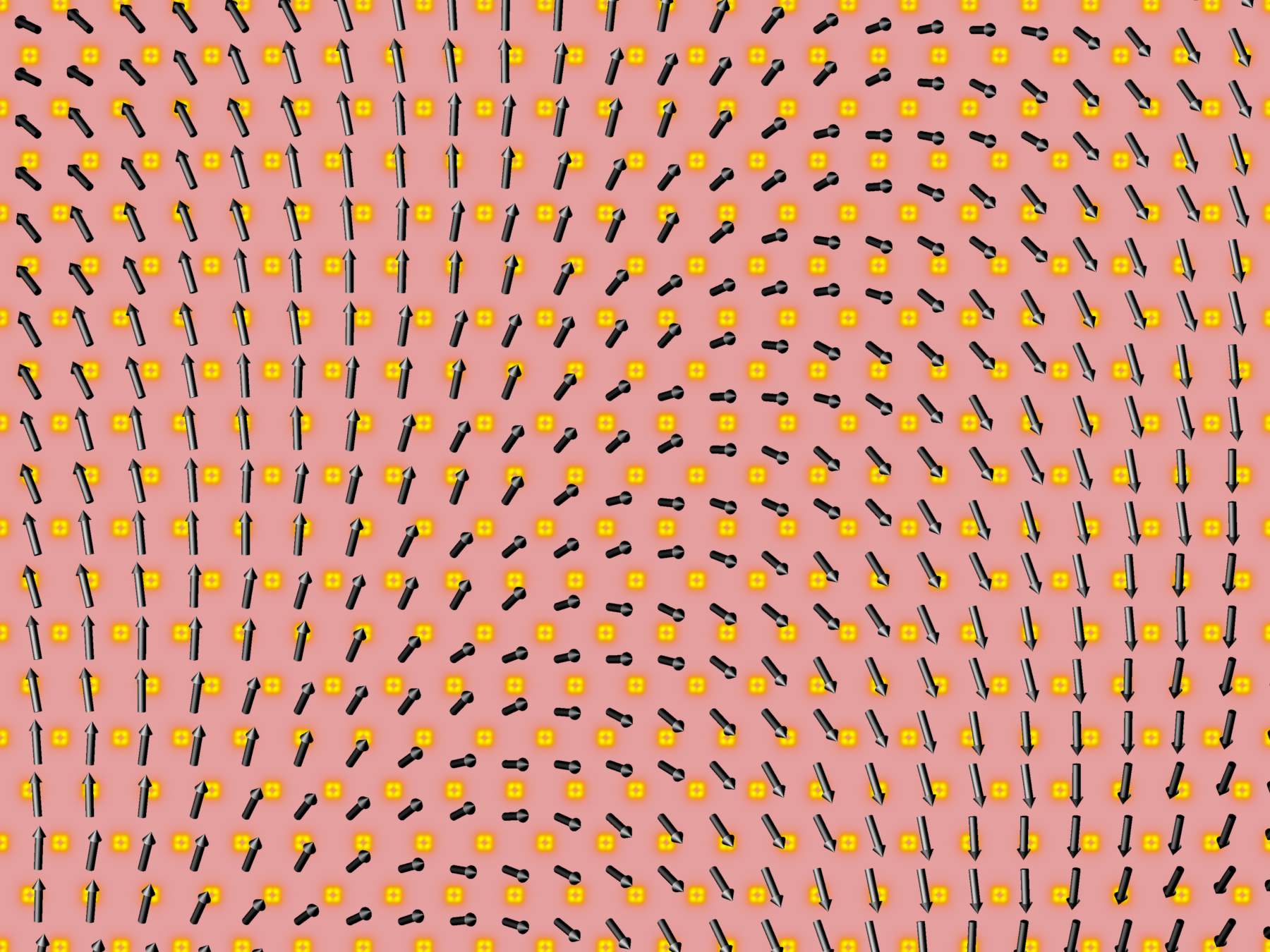}
 \caption{}
\end{subfigure}\quad
\begin{subfigure}{0.48\columnwidth}
 \includegraphics[width=\linewidth]{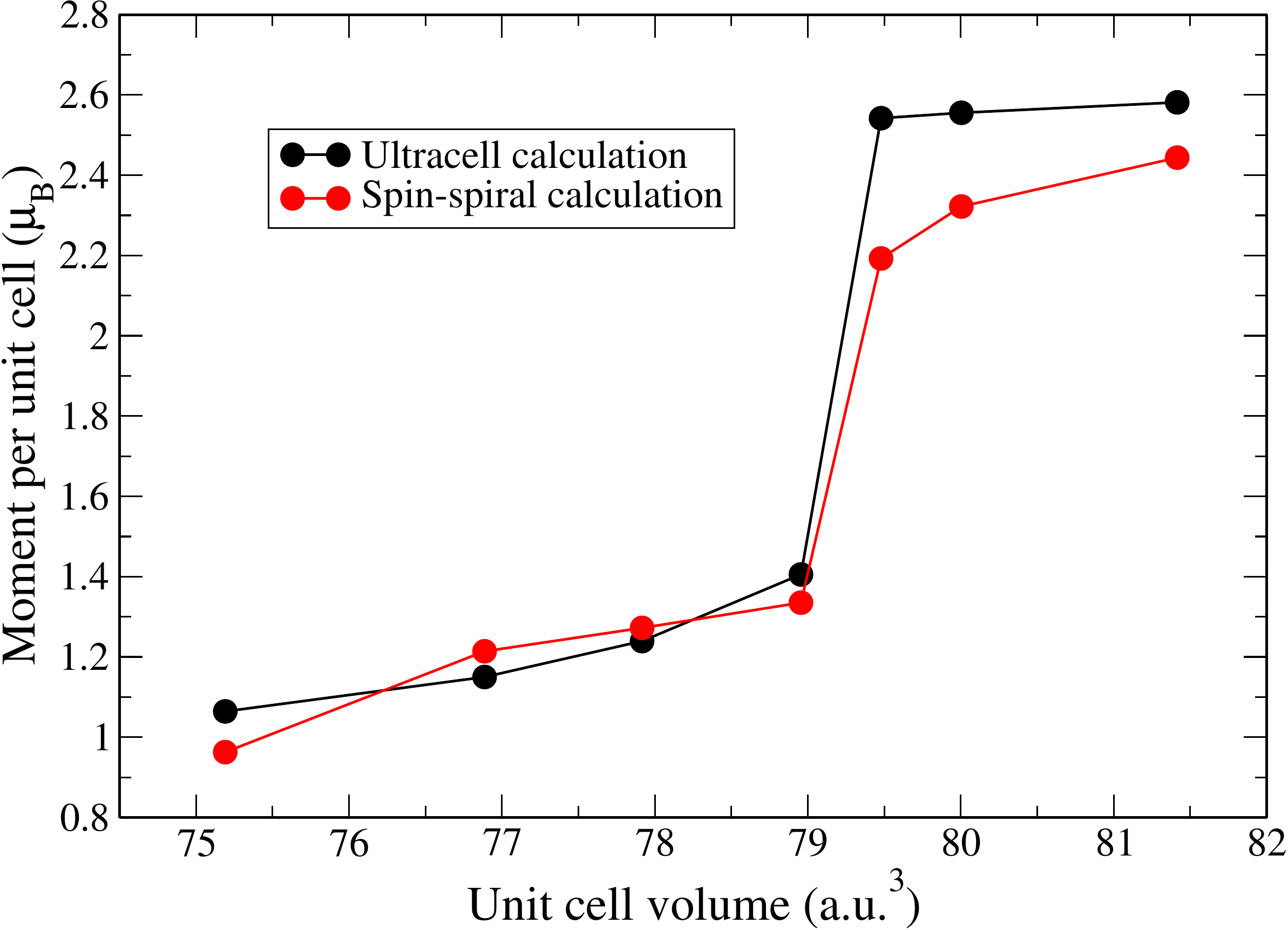}
 \caption{}
\end{subfigure}
\caption{\raggedright (a) Ultra long-range magnetization density of
 $\gamma$-Fe plotted in the plane perpendicular to $[0 0 1]$.
 The color indicates the magnitude of the magnetization and the
 arrows indicate direction. The modulation encompasses 32 unit cells in the
 $[1 0 0]$ direction. (b) Plot of moment against unit cell
 volume for both the long-range and spin-spiral ansatz.}\label{fig:Fe}
\end{figure}
The first numerical test deals with the so called $\gamma$ phase of Fe. Previous calculations\cite{PhysRevB.66.014447} have shown that the spin-spiral state has the lowest energy compared to several commensurate ferromagnetic and anti-ferromagnetic structures. The ultra long-range method allows us to address the question whether the much larger variation freedom associated with ultra-cell still yields the spin-spiral as ground state. We performed a traditional spin-spiral calculation and an ultra-cell calculation for this materials. The parameters used are as follows:
Ultracell ${\bf k}$-point grid:
$1\times 12\times 12$, ${\bf Q}$-point grid: $32\times 1\times 1$,
ultracell: $32\times 1\times 1$ unit cells. A single unit cell was used for
the spin-spiral calculation with a $12\times 12\times 12$ ${\bf k}$-point
grid and a ${\bf Q}$-vector of $1/32$.

An initial magnetic field is required to break the spin symmetry. To ensure an
unbiased calculation, we applied a random field to the ultracell calculation
and subsequently reduced it to zero.
Throughout the calculation, we enforced the constraint
$\int_{\rm unit} d^3 r \,{\bf m}_{{\bf Q}=0}({\bf r}) = 0$.
This ensures that the system is not drawn to a lattice-periodic
ferromagnetic solution. The magnetization converged
to an ordered state where the magnitude was constant over the
ultracell and only the direction varied (Fig. \ref{fig:Fe}(a)).
This corresponds precisely to the
spin-spiral state i.e. the ultra-cell calculations shows that the spin-spiral state is still the lowest energy solution. The overall magnitude of the magnetization is sensitive
to the lattice parameter and undergoes a transition from $\sim 1\mu_B$ to
$\sim 2.5\mu_B$ for this relatively small ${\bf Q}$-vector. As may be seen
in Fig. \ref{fig:Fe}(b), this behavior is observed for both the
ultracell and spin-spiral calculations.

\subsection{Spin density wave in bcc Cr}

\begin{figure}[ht]
\centering
\begin{subfigure}{0.48\columnwidth}
 \includegraphics[width=\linewidth]{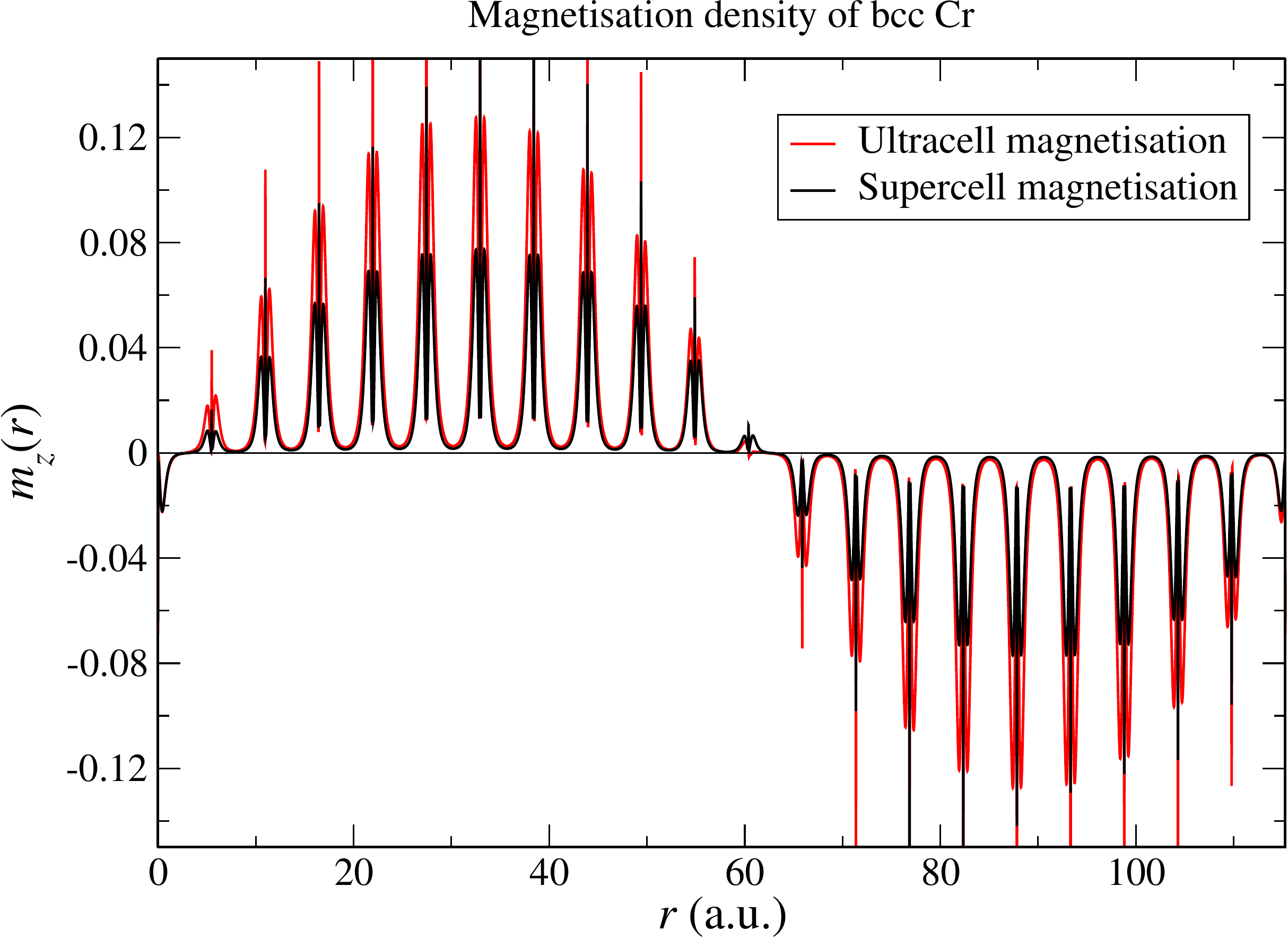}
 \caption{}
\end{subfigure}\quad
\begin{subfigure}{0.48\columnwidth}
 \includegraphics[width=\linewidth]{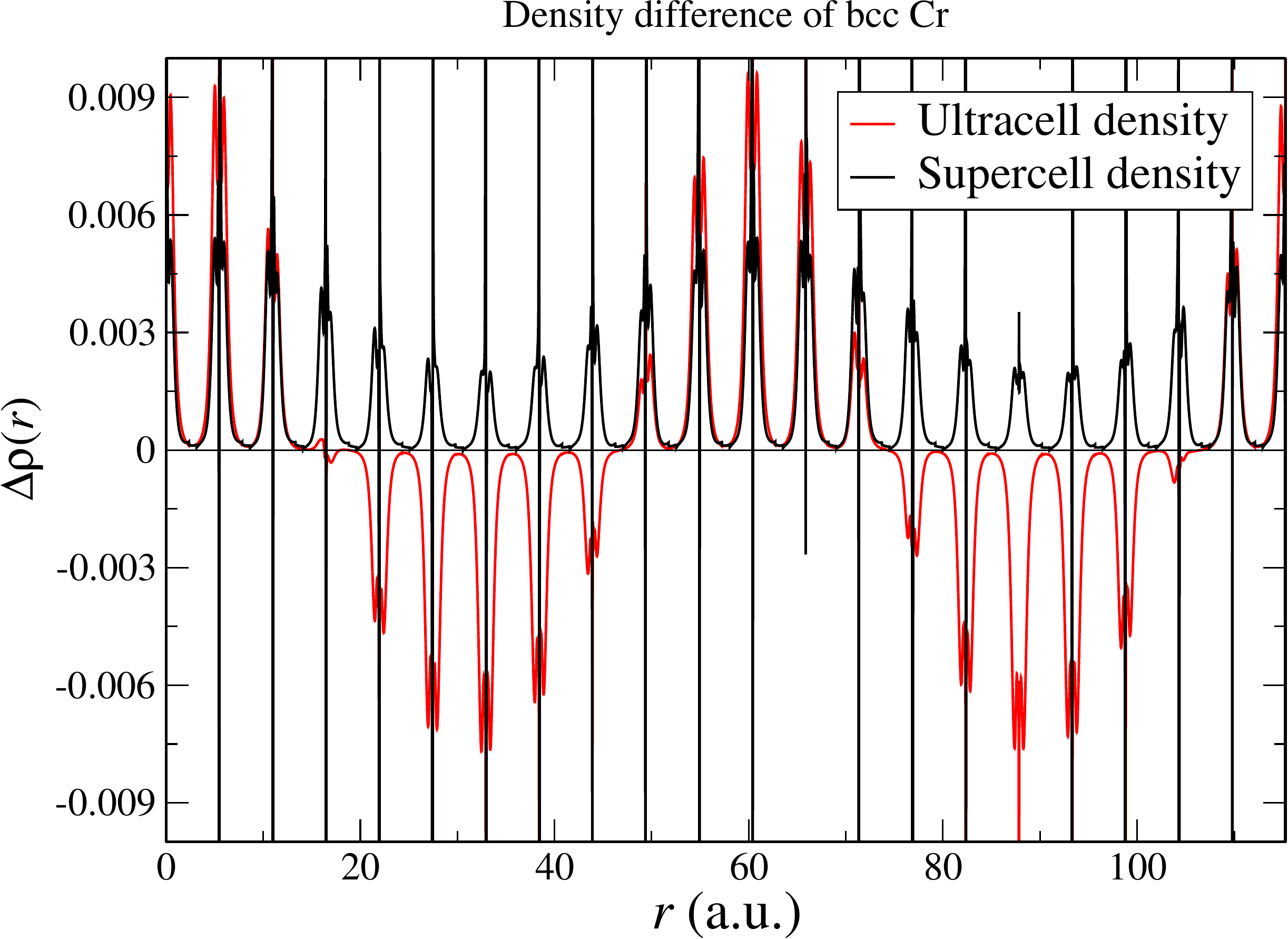}
 \caption{}
\end{subfigure}
\caption{\raggedright (a) Magnetization density for bcc Cr over 21 unit cells.
(b) Change in density over the same range. For the ultracell, this was
generated by setting $\rho_{{\bf Q}=0}({\bf r})$ in Eq. (\ref{eq:coeff}) to zero.
For the super-cell, the lattice-periodic density was subtracted leaving just
the modulated density.}\label{fig:Cr}
\end{figure}

In a second test, we aim at calculating the spin density wave (SDW) state in Cr.
The existence of a SDW in Cr is well known and the first research dates back to
around 1960 \cite{corliss_1959,bykov_1960,overhauser_1962}. Despite this,
computing the SDW state within DFT remains difficult and has been the topic of
many studies \cite{moruzzi_1978,kubler_1980,skriver_1981,kulikov_1982,
kulikov_1987,chen_1988,moruzzi_1992,singh_1992,hirai_1997,
guo_2000,bihlmayer_2000,schafer_2000,hafner_2001}, with partially conflicting
results \cite{cottenier_2002}. It is likely that a SDW is not the true ground
state of Cr within DFT \cite{hafner_2002}, however we will not focus here on
the inherent complexities of the system.
This state is not achievable by the spin-spiral ansatz because the
magnitude of the moment changes but not its direction, thus a super-cell
calculation is required. Cr is an excellent test scenario, as the periodicity
of the SDW is $\sim 20.83$ unit cells, which is still well within computational
reach of the super-cell approach.

For our comparison, we use the LSDA and a
lattice parameter of \SI{2.905}{\angstrom} as suggested by Cottenier
et al.\cite{cottenier_2002}. 
We consider $21 \times 1 \times 1$ unit cells of bulk Cr for both
super-cell and ultracell calculations.
For the super-cell we used a $1 \times 12 \times 12$ ${\bf k}$-point grid. A
randomized symmetry breaking magnetic field was used to start the calculation and
subsequently reduced to zero. Spin-orbit coupling is also included.
Our super-cell calculation reproduces the result by
Cottenier {\it et al.}\cite{cottenier_2002}.
For the ultracell we also used a $1 \times 12 \times 12$ ${\bf k}$-point grid
with a $21 \times 1 \times 1$ ${\bf Q}$-points grid corresponding to
a grid of $11 \times 1 \times 1$ $\bm{\kappa}$-points to obtain the best
possible sampling of the xc-potential and -field. Around 60 empty states in
the lattice-periodic basis are used to provide enough degrees of freedom during
the convergence. We started with a randomized initial field that was reduced after
each step. Throughout the calculation, we enforced the constraint
$\int_{\rm MT} d^3 r \,{\bf m}_{{\bf Q}=0}({\bf r}) = 0$ for each muffin-tin.
This ensures that the system is not drawn to a lattice-periodic
anti-ferromagnetic solution.

Our results are shown in Fig. \ref{fig:Cr}. Specifically, Fig. \ref{fig:Cr}(a)
shows the comparison of the magnetization in the SDW state, as obtained from
the super-cell and ultracell calculations. The maximum moment of the
ultracell calculations is larger than that of the super-cell, 1.174 $\mu_B$ and
0.712 $\mu_B$, respectively. This we attribute to the fact that the ultra
long-range calculation is performed in the basis of Kohn-Sham states and not
in the original LAPW basis for which the linearization energies are optimally
adjusted. It is also known that LSDA calculations this of system are particularly
sensitive to the basis and the moment depends strongly on the
lattice parameter\cite{cottenier_2002}.

In Fig. \ref{fig:Cr}(b) we present the charge density wave (CDW) which is known
to stabilize alongside the SDW with twice the period. While obtaining the CDW in
the ultracell is straight-forward (as all $\rho_{\bf Q}({\bf r})$ are known),
it is numerically more challenging to extract it for the super-cell. We did this
by subtracting the density from the calculation of a single unit cell.
We obtain the same periodicity in both calculations as well as a comparable magnitude.

\subsection{Saw-tooth potential in Si}

\begin{figure}[ht]
\centering
\begin{subfigure}{0.48\columnwidth}
 \includegraphics[width=\linewidth]{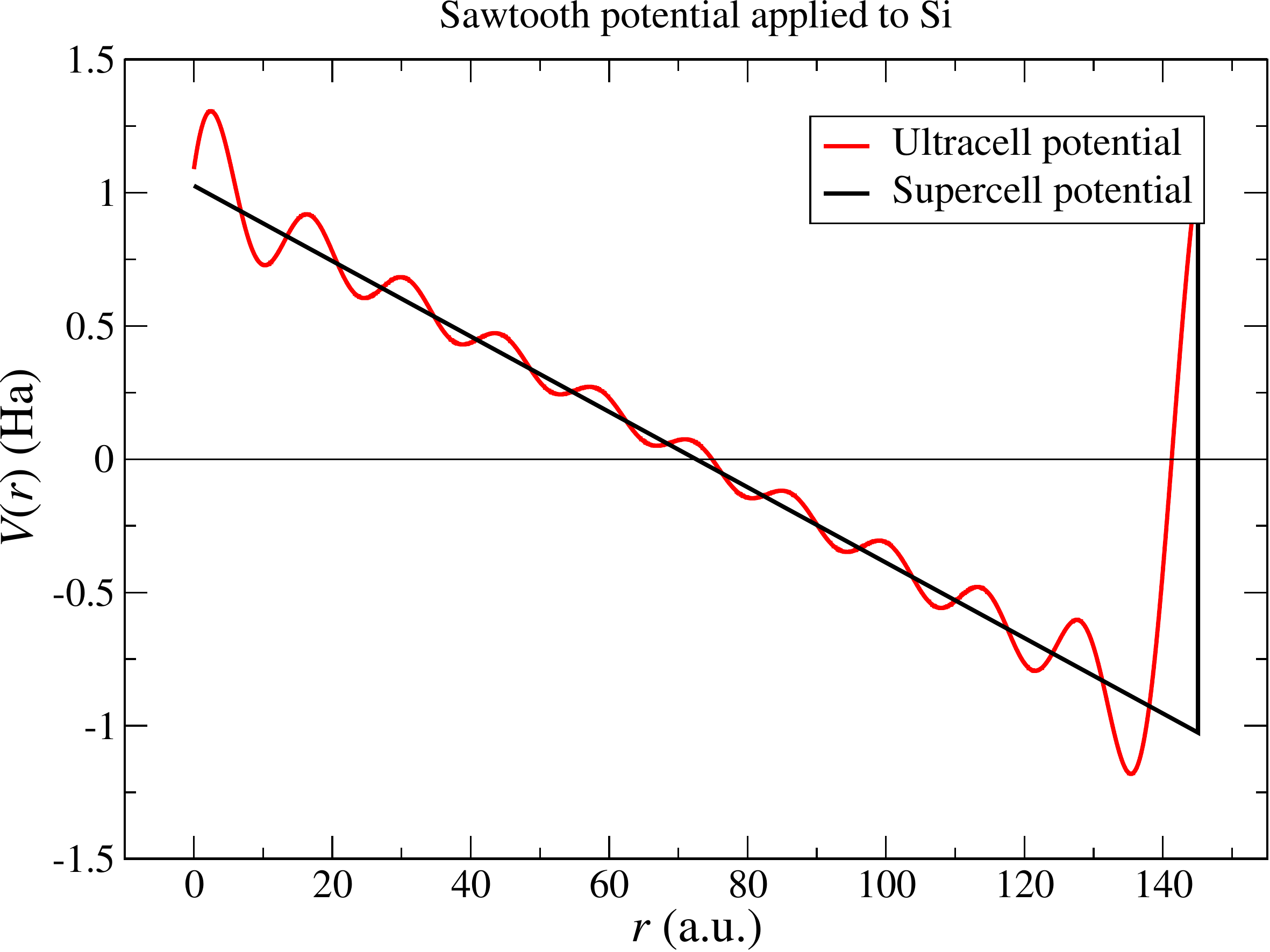}
 \caption{}
\end{subfigure}\quad
\begin{subfigure}{0.48\columnwidth}
 \includegraphics[width=\linewidth]{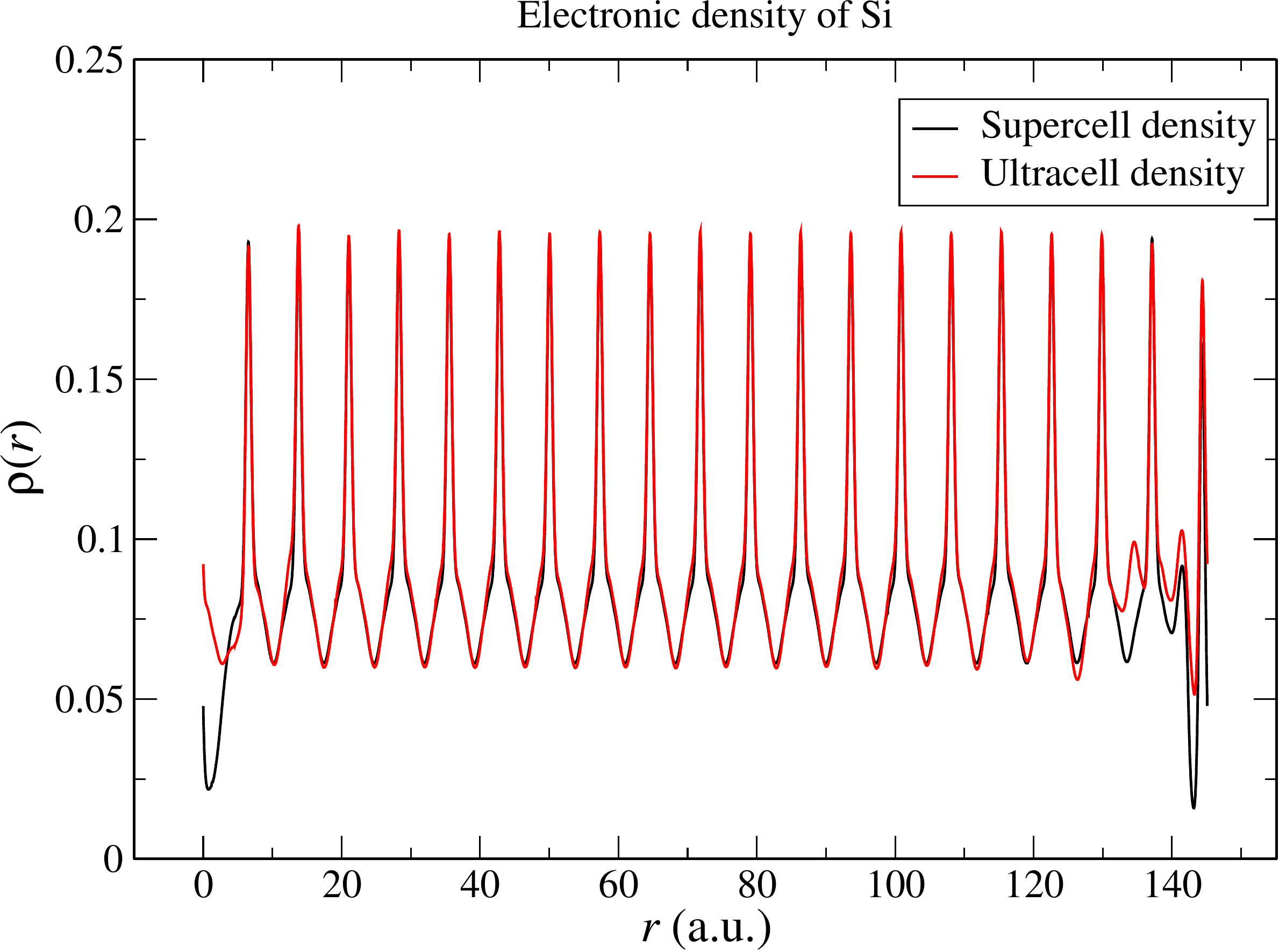}
 \caption{}
\end{subfigure}
\caption{\raggedright (a) Saw-tooth potential applied to 20 unit cells of
 silicon in the ultracell and super-cell. The ultracell potential is
 smoother than that of the super-cell because of the relatively small number
 of ${\bf Q}$ vectors used to expand it. (b) The resultant ground state density
 in the ultracell and super-cell. This was plotted along a line which was slightly
 off-set from the atomic centers in order to avoid the very high densities near
 the nuclei.}\label{fig:Si}
\end{figure}

The previous two examples involved modulations in long-range magnetic order,
but we also need to test the method with long-range, external electrostatic fields.
This is in anticipation of a future development where ultra long-range TDDFT
calculations are performed in conjunction with Maxwell's equations. In such a
scenario, an electromagnetic wave propagating through the solid could have a
periodicity of many hundreds of unit cells. The long-range ansatz should be
ideal for performing such a simulation.
With that goal in mind, we apply a simple saw-tooth potential to
silicon over a range of 20 unit cells to check if the ultracell calculation
agrees with its super-cell equivalent. This corresponds to a constant
electric field, at least near the center of the saw-tooth.
Both calculations were performed with a $4\times 4\times 4$ ${\bf k}$-point
grid. The ultracell calculation used a ${\bf Q}$-point grid of $20\times 1\times 1$
with a basis of 60 empty states per ${\bf k}$-point.
The applied electric field was $0.01$ in atomic units
and the corresponding saw-tooth potential is plotted in Fig. \ref{fig:Si}(a).
As can been seen, there is a difference between the ultracell and super-cell
potentials. The ultracell potential is expanded in a finite set of ${\bf Q}$
vectors and thus contains oscillations on the length scale of the longest vector.
The super-cell potential is a sharp saw-tooth. Despite this difference, the
two densities plotted in Fig. \ref{fig:Si}(b) are broadly the same with strong
screening near the center and charge accumulation and depletion near the
edges. We note that for the intended purpose of describing propagating
light through solids, the resulting electric and magnetic fields will be
well expanded with a finite number of ${\bf Q}$ vectors.

\subsection{Long-range electrostatic potential in LiF}

\begin{figure}[ht]
 \includegraphics[width=0.9\linewidth]{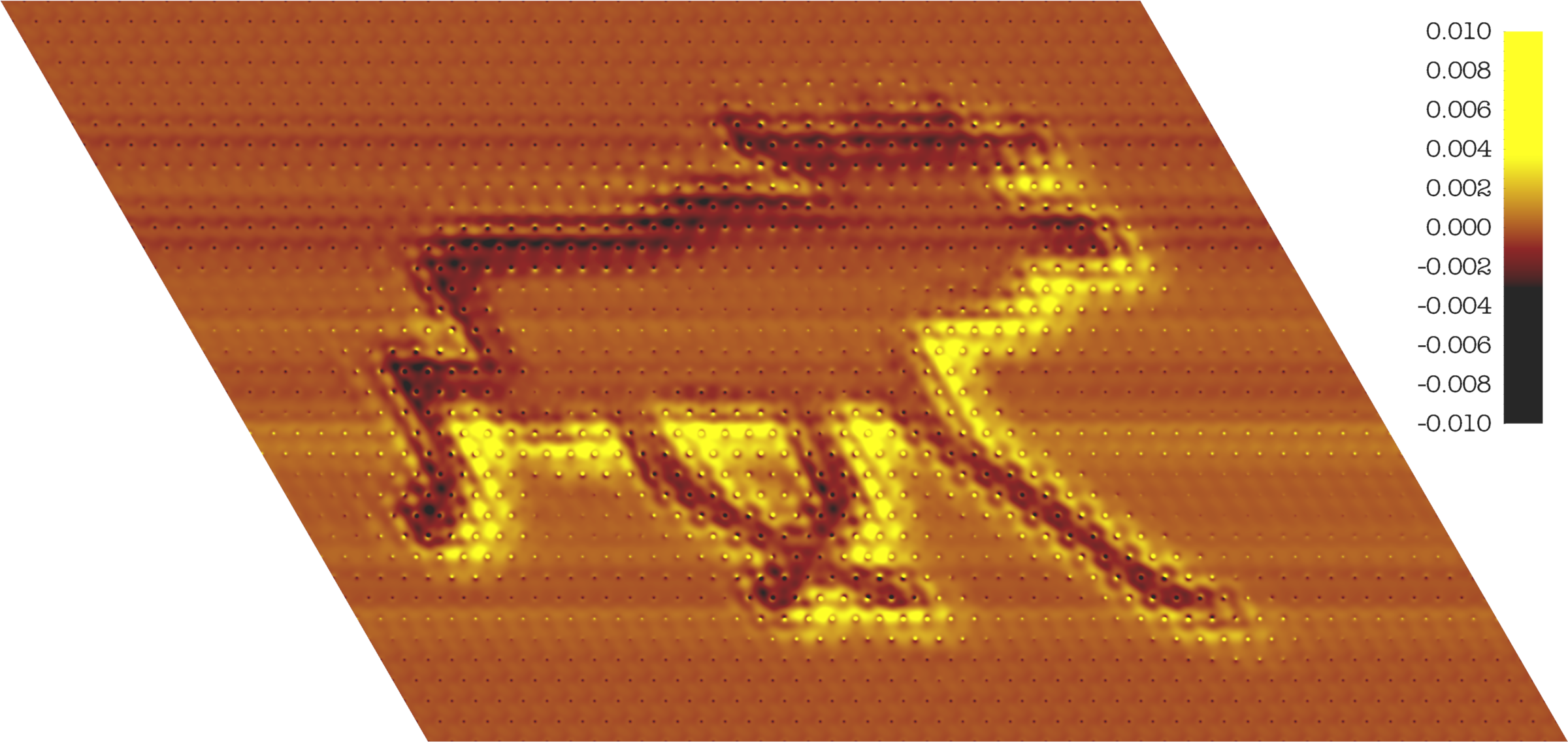}
\caption{\raggedright Self-consistent density without the
 $\rho_{{\bf Q}=0}({\bf r})$ term for a 3456 atom ultracell of
 LiF with an artificial external potential. The plotting plane is perpendicular
 to $[0 0 1]$ and contains $48 \times 36$ unit cells.}\label{fig:LiF}
\end{figure}

Lastly we perform a calculation which is too large for a super-cell.
Rather than attempting to model a physical phenomenon at this stage, we
simply apply an arbitrarily chosen electrostatic potential to an insulator,
in this case LiF.
An ultracell of $48\times 36\times 1$ unit cells was constructed with
an equivalent ${\bf Q}$-point grid. The number of empty states per
${\bf k}$-point was taken to be 4 to keep the memory requirements to within
those of our computer.
The resultant change in density away from unit cell periodicity
is plotted in Fig. \ref{fig:LiF}. As the potential is artificial, the important
metric here is the computational effort expended in reaching the
self-consistent solution.
The rate of convergence is fairly slow because of the effect of the long-range
Coulomb interaction, and thus we performed 170 iterations of the
self-consistent loop. The calculation was performed on 480 CPU cores and
each iteration took about 40 minutes. This level of performance for an
all-electron calculation indicates that physical phenomena involving
modulations of the electronic state
over hundreds or thousands of unit cells are within reach of this approach.

\section{Conclusion and outlook}
We have developed a method which makes possible the {\it ab-initio} treatment
of hitherto uncomputable length-scales in solids. This consists of a
modified Bloch ansatz and a set of Kohn-Sham equations
which have to be solved self-consistently. The underlying lattice of nuclear
charges is still periodic on the unit cell length scale but the electronic state can
accommodate arbitrary modulations on any length scale.
Based on our experience with the all-electron Elk
code, we are confident that this method can be efficiently implemented in most
existing solid-state electronic structure codes.
We demonstrated the capabilities of the novel method by
solving an arbitrary external potential applied to nearly 3500 atoms
of LiF. Additionally, we showed that our method can reproduce the results obtained by
super-cell calculations on smaller length scales for both insulators and
magnetic solids.
The method presented in this paper opens up exciting possibilities of future
research: on the technical level, a derivation of long-range and explicitly
${\bf Q}$-dependent xc-potentials
(see, for example, Pellegrini {\it et al.} \cite{PhysRevB.101.144401}).
On the applied level, our method could pave the way to calculations of mesoscopic
systems, such as magnetic domain walls or skyrmions, which have so far been out
of reach for {\it ab-initio} methods like DFT. 
Furthermore, the novel technique is straightforwardly incorporated in real-time TDDFT calculations which, when combined with the solution of Maxwell's equations, will give access to the propagation of electromagnetic radiation through extended solids within a genuine \emph{ab-initio} description.

Acknowledgments: SS and TM would like to thank DFG for funding though QUTIF project. EKUG acknowledges financial support by European Research Council Advanced Grant Fact (ERC-2017-AdG-788890).

%\bibliography{bibliography}
%merlin.mbs apsrev4-1.bst 2010-07-25 4.21a (PWD, AO, DPC) hacked
%Control: key (0)
%Control: author (8) initials jnrlst
%Control: editor formatted (1) identically to author
%Control: production of article title (-1) disabled
%Control: page (0) single
%Control: year (1) truncated
%Control: production of eprint (0) enabled
%

\end{document}